\documentclass[aps,prl,twocolumn,groupedaddress,showpacs]{revtex4}
\usepackage{epsfig,amsmath}
\pdfoutput=1
\usepackage{color}
\definecolor{light-gray}{gray}{0.5}
\usepackage[colorlinks=true,pagebackref=true]{hyperref}
\hypersetup{
    pdftitle={A Novel Spin-Orbit Torque due to Conduction Electrons},
    pdfauthor={Ioan Tudosa},
    pdfsubject={},
    pdfkeywords={magnetism magnetization switching magneto-transport},
    citecolor=light-gray,
    linkcolor=light-gray,
    urlcolor=light-gray
    }

\begin{document}

\title{A Novel Spin-Orbit Torque due to Conduction Electrons}

\author{Ioan Tudosa}
\email[Electronic address:]{itudosa@gmail.com}
\noaffiliation

\date{\today}

\begin{abstract}
The anomalous Hall effect is mainly used to probe the magnetization orientation in ferromagnetic materials. A less explored aspect is the torque acting back on magnetization, an effect that can be important at high currents. The spin-orbit coupling of the conduction electrons causes spin-up and spin-down electrons to scatter to opposite sides when a charge current flows in the sample. This is equivalent to a spin current with orientation and flow perpendicular to driving charge current, leading to a non-equilibrium spin accumulation that exerts a torque on bulk magnetization through the s-d exchange interaction. The symmetry of this toque is that of an uniaxial anisotropy along the driving current. The large screening currents generated with laser pulses in all-optical magnetic switching experiments make for practical uses of this torque.
\end{abstract}

\pacs{75.47.-m,75.47.Np,73.50.Jt,72.15.Gd,72.25.-b,75.50.Bb,73.43.Qt,75.78.Jp,75.78.-n,72.25.Ba}

\maketitle

\section{Introduction}

The spin Hall effect is often used to generate spin currents that exert torques in heterostructures made of non-magnetic and ferromagnetic layers \cite{Garello:2013}. Non-magnetic high Z materials are used to get a sizable spin-orbit coupling that generates a large spin current. The common 3d ferromagnetic materials Fe, Co, Ni have low Z number and moreover the exchange interaction dominates spin-orbit coupling. The most used scheme is to generate a spin current in the high Z non-magnetic material and transport it to the ferromagnetic material. There is however an anomalous Hall effect \cite{Nagaosa:2010} ferromagnetic materials and it relies on a bulk spin-orbit coupling present in ferromagnetic materials, not as big as in high Z materials but still important. For high enough currents this spin-orbit coupling can exert a significant influence on magnetization dynamics as shown recently in pure iron and iron-cobalt alloys \cite{Pattern:2017}.

When there is spin-orbit coupling and scattering, the spin-up and spin-down electrons (with quantization axis perpendicular to the their motion direction) will be deflected and separated to opposite sides as shown in Fig.~\ref{AHEcurrent}. This separation is equivalent to generating a spin current and producing a non-equilibrium spin accumulation that exerts a torque on magnetization through s-d exchange interaction. Surprisingly, the sign of such a torque does not depend on the sign of the current, as if the current imposes an uniaxial anisotropy axis. Controlling the anisotropy temporarily with a current may be used in magnetization switching and an easy method to generate high currents is by screening strong lases pulses in metallic ferromagnetic samples. This spin-orbit torque has been overlooked in all-optical magnetic switching experiments \cite{Stanciu:2007,Ostler:2012}.

\section{Spin Current in Anomalous Hall Effect}

\begin{figure}
\includegraphics[]{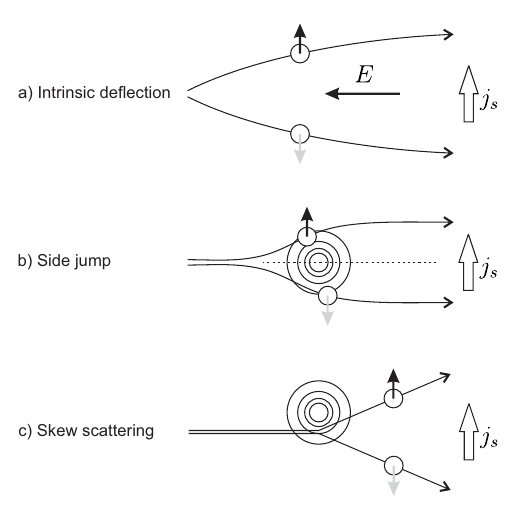}
\caption{\label{AHEcurrent} There are three main mechanisms of spin-dependent deflection of conduction electrons (open circles) involved in anomalous Hall Effect. The spin-orbit coupling acts only on the spin component perpendicular to the motion direction of electrons. It is along this perpendicular direction that we define our quantization axis for spin-up and spin-down (arrows on open circles). The deflection of spins is equivalent to a spin current perpendicular to the main motion of the charge. (a) Intrinsic deflection arises from the electric field induced interband coherence leading to a perpendicular velocity contribution related to the Berry phase curvature of band structure \cite{Nagaosa:2010}. (b) Side jump occurs because velocity is deflected in opposite directions upon entering and leaving an impurity. (c) Skew scattering is an asymmetric scattering of the spins due to spin-orbit coupling.}
\end{figure}

\begin{figure}
\includegraphics[]{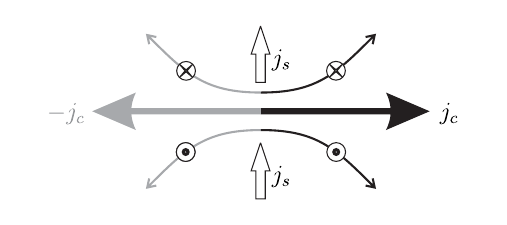}
\caption{\label{Scurrent} Due to spin-orbit coupling electrons get
deflected to one side depending on their spin, shown here as in or
out of plane of the picture. On changing the charge current
direction (or the sign of electron velocity) the trajectory shape
and curvature does not change and the spins with the same
direction get deflected to the same side. Thus the spin current
sign does not depend on the sign of the driving charge current.
The charge current effectively induces an anisotropy axis for
spins.}
\end{figure}

At the root of anomalous Hall effect (AHE) is a deflection of charge carriers due to the spin-orbit coupling as they travel under a driving current. There are three main mechanism for deflection shown in Fig.~\ref{AHEcurrent}, adapted from reference \cite{Nagaosa:2010}. The first mechanism is intrinsic as the electric field producing the driving current induces interband coherence and electrons get a perpendicular velocity related to the Berry phase curvature of their band structure. The other two mechanism are related to spin-dependent scattering on impurities. In a side jump electrons get opposite deflection upon entering and leaving the impurity while skew scattering is due to asymmetric scattering of spin-up and spin-down. Regardless of the specific mechanism, the deflection is equivalent to a spin current because opposite spins move in opposite directions. This spin current has the same direction no matter the sign of the driving charge current, as illustrated in Fig.~\ref{Scurrent}. The curvature of the deflection does not depend on whether the electron is entering or leaving the area.

The Hall effect resistivity \(\rho_{\perp}\) includes a standard component proportional to the magnetic field (\(\propto r_0 H\)) and an anomalous component proportional to magnetization (\(\propto r_a M\)), as shown in Eq.~\ref{eq:Hallresist} where \(r_0\) and \(r_a\) are coefficients that describe the strength of the respective effects. Both magnetic field and magnetization need to have components perpendicular to the current direction; the parallel components cause no effect. Anomalous Hall effect can be 1-2 orders of magnitude higher than the standard Hall effect in ferromagnetic materials.

\begin{equation}
 \label{eq:Hallresist}
 \rho_{\perp}=r_0 H+r_a M
\end{equation}

In terms of conductivities, the anomalous Hall conductivity \cite{Nagaosa:2010} at magnetic saturation \(\sigma_{\perp}\) is around 1\% of the standard longitudinal conductivity \(\sigma_{\parallel}\) involved in driving the charge current \(j_c\). The anomalous Hall angle \(\theta_H\) between the driving electric field and the total current is given by Eq.~\ref{eq:Hallangle}.

\begin{equation}
 \label{eq:Hallangle}
 \theta_H \approx \tan \theta_H=\frac{\sigma_{\perp}}{\sigma_{\parallel}}
\end{equation}

In the absence of charge accumulation at the edges, the perpendicular charge current \(j_{\perp}\) due to anomalous Hall effect deflection can be calculated by Eq.~\ref{eq:Hallcurrent}.

\begin{equation}
 \label{eq:Hallcurrent}
 j_{\perp}=j_c\tan\theta_H=\frac{\sigma_{\perp}}{\sigma_{\parallel}}j_c
\end{equation}

Charge carriers in ferromagnetic materials have a spin asymmetry \(P\), i.e. there are unequal numbers of spin-up and spin-down carriers. Thus the perpendicular anomalous Hall current carries also a spin, with the magnitude of the spin current shown in Eq.~\ref{eq:Hallspincurrent}.

\begin{equation}
 \label{eq:Hallspincurrent}
 j_s=\frac{1}{P}\frac{\hbar/2}{e}j_{\perp}=\frac{\hbar}{2 e P}\frac{\sigma_{\perp}}{\sigma_{\parallel}}j_c
\end{equation}

\begin{figure}
\includegraphics[]{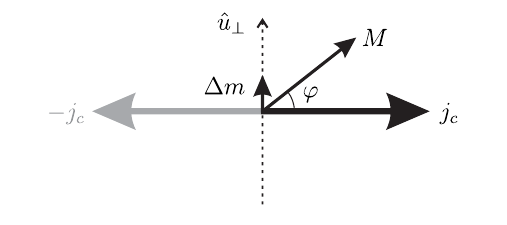}
\caption{\label{Orientation} Relative orientation of driving charge current \(j_c\), magnetization \(M\), perpendicular quantization axis \(u_{\perp}\) and non-equilibrium magnetic moment density \(\Delta m\) generated by deflection of electrons due to spin-orbit coupling.}
\end{figure}

While all deflected spin-up and spin-down carriers contribute to the spin current, only their spin asymmetry contributes to the perpendicular charge current. The spins in excess will carry the charge current toward their respective side. The quantization axis for spin-up and spin-down is perpendicular to the driving charge current and if magnetization is not aligned with this axis \(\hat{u}_{\perp}\) then a projected component must be factored into Eq.~\ref{eq:Hallspincurrent}, as in Eq.~\ref{eq:HallspincurrentM}. With \(\varphi\)  being the angle between magnetization and driving current the projection is \(\hat{u}_{\perp}\cdot\hat{M}=\sin\varphi\) as illustrated in Fig.~\ref{Orientation}.

\begin{equation}
 \label{eq:HallspincurrentM}
 j_s=\frac{\hbar}{2 e P}\frac{\sigma_{\perp}}{\sigma_{\parallel}} (\hat{u}_{\perp}\cdot\hat{M}) j_c
\end{equation}

\section{Torque from Nonequilibrium Spin Accumulation}

The spin current in Eq.~\ref{eq:HallspincurrentM} leads to a non-equilibrium excess of spin whose volume density \(\Delta s_{\perp}\) can be inferred as in Eq.~\ref{eq:Spindensity} since the charge and spin carriers move roughly with Fermi velocity \(v_F\). The transverse spin decoherence length is very short, on the order of Angstroms \cite{Manchon:2012}, so the spin current is very efficiently converted to nonequilibrium spin accumulation.

\begin{equation}
 \label{eq:Spindensity}
 \Delta s_{\perp}\approx\frac{j_s}{v_F}
\end{equation}

This excess of conduction electron spin \(\Delta s_{\perp}\) exerts a torque on the bulk magnetization spin \(S\) via the s-d exchange interaction that has a Hamiltonian \(H_{sd}=-J_{sd}\vec{S}\cdot\Delta\vec{s}_{\perp}\), where \(J_{sd}\) is the s-d exchange energy.

The expression for the torque acting on the bulk magnetization spin comes from the Ehrenfest theorem applied to its evolution under the s-d exchange interaction \cite{Manchon:2012}, resulting in a precessional dynamics.

\begin{equation}
 \label{eq:TorqueEhrenfest}
 \frac{\partial \vec{S}}{\partial t}=\frac{J_{sd}}{\hbar} \vec{S} \times \Delta \vec{s}_{\perp}
\end{equation}

With the equalities \({\vec{M}}/{M_s}=-{\vec{S}}/{S}\), \(\Delta \vec{m}=-\gamma \Delta\vec{s}_{\perp}\) and multiplying Eq.~\ref{eq:TorqueEhrenfest} by gyromagnetic ratio \(\gamma=g\mu_B/\hbar\), we obtain the equation \ref{eq:TorqueEhrenfestM}

\begin{equation}
 \label{eq:TorqueEhrenfestM}
 \frac{\partial \vec{M}}{\partial t}=-\frac{J_{sd}S}{\hbar M_s} \vec{M} \times \Delta \vec{m}
\end{equation}

Such a torque can be added to the standard Landau-Lifshitz-Gilbert equation Eq.~\ref{eq:LLG}, where we have introduced the non-equilibrium magnetic moment density \(\Delta \vec{m}\),  external magnetic field \(\vec{H}\) and damping coefficient \(\alpha\) of magnetization dynamics.

\begin{equation}
 \label{eq:LLG}
 \frac{d\vec{M}}{dt} = -\gamma\vec{M}\times\vec{H} + \frac{\alpha}{M_s}\vec{M}\times\frac{d\vec{M}}{dt} - \frac{J_{sd}S}{\hbar M_s}\vec{M}\times\Delta\vec{m}
\end{equation}

The non-equilibrium spin density acts as an effective magnetic field \(H_{SO}\) with the amplitude given by Eq.~\ref{eq:Hso}. Only a small fraction of the spins compared to bulk magnetization is needed because the exchange interaction is pretty strong (on the order of 0.7 eV for iron \cite{Mamy:2002}). As long as there is a spin-orbit coupling and a current is flowing, there will be a torque acting on magnetization.

\begin{equation}
 \label{eq:Hso}
 H_{SO}=\frac{J_{sd}S}{g\mu_B}\frac{\Delta m}{M_s}
\end{equation}

Using Eqs.~\ref{eq:HallspincurrentM} and \ref{eq:Spindensity} we arrive at an expression of effective spin-orbit field in terms of the driving charge current, in Eq.~\ref{eq:Hsocurrent}. This effective field is along the axis \(\hat{u}_{\perp}\), in a plane perpendicular to the charge current.

\begin{equation}
 \label{eq:Hsocurrent}
 H_{SO}=\frac{J_{sd}S}{M_s} \frac{1}{2 e P v_F} \frac{\sigma_{\perp}}{\sigma_{\parallel}} (\hat{u}_{\perp}\cdot\hat{M}) j_c
\end{equation}

The spin accumulation in the conduction electrons is felt with little delay as a torque on the bulk magnetization due to relatively strong s-d exchange interaction.

\section{Symmetry of Torque}

The angle between magnetization \(\vec{M}\) and driving charge current \(\vec{j}_c\) determines the magnitude of spin-orbit torque in Eq.~\ref{eq:TorqueEhrenfestM}. Using Eq.~\ref{eq:Hsocurrent} we arrive at a \(\sin 2\varphi\) dependence in Eq.~\ref{eq:Tso}, meaning a maximum torque occurs when magnetization makes a 45 degree angle with the driving current. This is analogous to the angle bias between current and magnetization to achieve best sensitivity in magnetic sensors based on anisotropic magnetoresistance effect (where spin-orbit coupling also plays a critical role).

\begin{equation}
 \label{eq:Tso}
 \begin{aligned}
 T_{SO} &= \vec{H}_{SO}\times\vec{M} \\
 &\propto (\hat{u}_{\perp}\cdot\hat{M}) (\hat{u}_{\perp}\times\hat{M}) = \sin\varphi \cos\varphi \propto\sin 2\varphi
 \end{aligned}
\end{equation}

\begin{figure}
\includegraphics[]{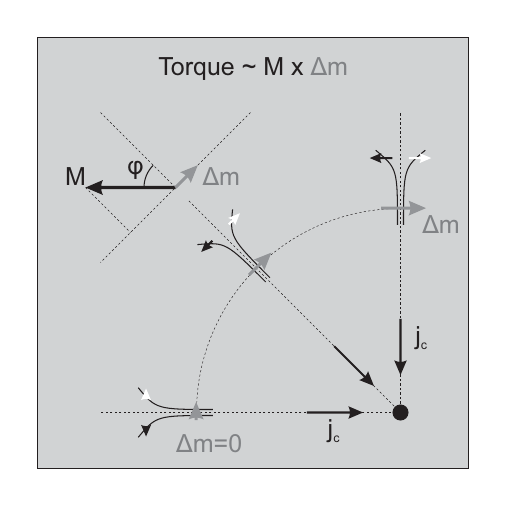}
\caption{\label{SOtorquesM} Illustration of three different angles
between driving charge current \(j_c\) and magnetization \(M\).
The torque is dependent on both the size and orientation of
non-equilibrium magnetic moment density \(\Delta m\) generated by
deflection due to spin-orbit coupling. The up/down deflected spins
(black/white color) are perpendicular to the sample plane for the
shown trajectories. In parallel configuration \(\Delta m\) is
zero, while in perpendicular configuration \(\Delta m\) is maximum
but its orientation is along \(M\) resulting in zero torque too.
Maximum torque is achieved at 45 degrees.}
\end{figure}

An illustration of the Eq.~\ref{eq:Tso} is in Fig.~\ref{SOtorquesM} where three different orientations of current with respect to magnetization are shown. When the magnetization is parallel with the current there is no spin-orbit deflection of the spins (\(\hat{u}_{\perp}\cdot\hat{M}=0\)); there has to be a perpendicular spin component for the deflection to happen. When magnetization is perpendicular to the current, the non-equilibrium spin density is maximum but is also parallel with magnetization and the torque is zero (\(\hat{u}_{\perp}\times\hat{M}=0\)); the only effect is a slight increase or decrease in magnitude of magnetization but not its direction. Both these factors are simultaneously optimized at 45 degrees, hence this is the angle where the torque is maximum.

As mentioned in Fig.~\ref{Scurrent} the torque does not depend on the sign of the charge current so it is sufficient to show only a quadrant of the whole range of angles in Fig.~\ref{SOtorquesM}. The \(\sin 2\varphi\) angular dependence suggests that one can think of the current as imposing a uniaxial magnetic anisotropy along its direction. The torque being the angular derivative of an energy, we find that the anisotropy energy is proportional to \(\cos^2\varphi\), very much like a standard uniaxial anisotropy. An experiment that shows this torque symmetry graphically has been done at SLAC National Accelerator Laboratory \cite{Pattern:2017, Back:1998} where all angles are sampled simultaneously in a setup with radial current similar to Fig.~\ref{SOtorquesM} (like in a Corbino effect geometry).

\section{Applications Using Large Screening Currents of Laser Pulses}

\begin{figure}
\includegraphics[]{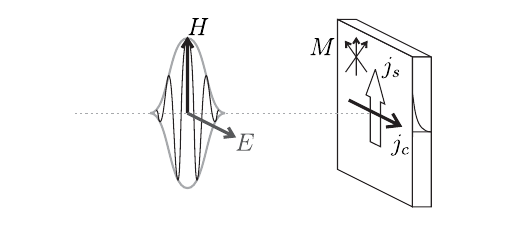}
\caption{\label{Laserpulse} Large charge currents can be generated by shining a strong and short laser pulse on a thin film of magnetic material. This picture illustrates the polarization and alternating nature of the magnetic field \(H\). The orientation of driving charge current \(j_c\) is in plane of the film and perpendicular to the magnetic field. The spin current \(j_s\) generated by deflection due to spin-orbit coupling exerts a torque on magnetization \(M\) and influences its switching behavior.}
\end{figure}

One method to generate large charge currents is through skin effect, a screening of a time varying magnetic field as illustrated in Fig.~\ref{Laserpulse}. Large field intensities can be achieved with a short femtosecond scale laser pulse, as used in all-optical magnetic switching \cite{Stanciu:2007,Ostler:2012}. Although the diffusion of an alternating magnetic field through a conductive sample follows an exponential decay profile with a characteristic skin depth \(\lambda\) \cite{Knoepfel:2000}, we can approximate the average current by assuming the magnetic field \(H\) is screened totally by the current density \(j_c\) inside the skin depth.

\begin{equation}
 \label{eq:Jcskin}
 j_c\approx\frac{H}{\lambda}
\end{equation}

At the visible and infrared wavelength the skin depth is on the order of 10-100 nanometer, while the peak magnetic field \(B=\mu_0 H\) of laser pulses used in optical magnetic switching can be in the range 0.1-1 Tesla. Thus the screening charge currents can be as high as \(10^{10} A/cm^2\) leading to a sizable spin-orbit torque.

Recent experiments in Fe and FeCo alloys \cite{Pattern:2017} have shown that this torque can be as high as 15\% that of the torque from the magnetic field but with a crucial difference. While an alternating magnetic field torque averages to zero, the spin-orbit torque does not change sign and has a net average value as in a rectifying effect. Using Eqs.~\ref{eq:Jcskin} and \ref{eq:LLG} a formula for the ratio between spin-orbit torque and magnetic field torque (first and last terms of Eq.~\ref{eq:LLG}) is shown in Eq.~\ref{eq:Kratio}.

\begin{equation}
 \label{eq:Kratio}
 \kappa = \frac{H_{SO}}{H}=\frac{J_{sd}S}{M_s} \frac{1}{2 e P v_F} \frac{\sigma_{\perp}}{\sigma_{\parallel}}\frac{1}{\lambda}
\end{equation}

Taking some typical values for iron \cite{Himpsel:1998,Mamy:2002,Nagaosa:2010} \(P=0.4,\, S=1,\, J_{sd}=0.7eV,\, v_F=0.5\times10^6m/s,\, \sigma_\perp/\sigma_\parallel=0.01,\, M_s=2.2T\) and \(\lambda=20nm\) we arrive at a ratio value of \(\kappa=0.4\). Lower saturation magnetization and short field pulses (small skin depth) make this torque ratio larger.

Laser pulses can also heat the electronic system bringing it close to Curie temperature and reducing the saturation magnetization \(M_s\). Since this quantity enters in the denominator of Eq.~\ref{eq:Kratio} the effect of the spin-orbit torque ratio will be enhanced. If spin polarization of conduction electrons \(P\) follows the bulk magnetization dynamics then its reduction also magnifies the torque ratio \(\kappa\). Large spin polarized currents generated by spin-orbit coupling have outsize effects on small magnetization.

\section{Conclusion}

The spin-orbit coupling that causes the anisotropic magnetoresistance and anomalous Hall effect in charge transport in magnetic materials also exerts a back reaction torque on the bulk magnetization. The torque has the symmetry of a uniaxial anisotropy with the axis along the charge current. The large currents needed for practical uses can be generated as screening eddy currents for the magnetic field in the short and strong laser pulses, such as those used in all-optical magnetic switching. The accompanying heat may bring the electronic system close to Curie temperature and enhance the effects of the torque.

\bibliography{SOtorque}

\end{document}